\documentclass[letter,10pt]{article} 
\usepackage{color,amssymb,amsthm,amsmath,amsxtra,fullpage,graphicx,hyperref} 

\usepackage{abstract}

\numberwithin{equation}{section}

\begin{document}

\def\bfS{{\textbf{S}}}
\def\mod{\text{ mod }}

\newcounter{construction-counter}
\setcounter{construction-counter}{0}

\def\const#1{%
\noindent\fbox{\parbox{\linewidth}{%
\addtocounter{construction-counter}{1}
{{\sc Construction \Roman{construction-counter}} \quad #1}}}}

\def\todo#1{\textcolor{red}{\textbf{**** TODO -- #1 ****}}}

\renewcommand{\qed}{\nobreak \ifvmode \relax \else
      \ifdim\lastskip<1.5em \hskip-\lastskip
      \hskip1.5em plus0em minus0.5em \fi \nobreak
      \vrule height0.75em width0.5em depth0.25em\fi}

\newtheorem{theorem}{Theorem}[section]
\newtheorem{lemma}[theorem]{Lemma}
\newtheorem{conjecture}[theorem]{Conjecture}
\newtheorem{proposition}[theorem]{Proposition}
\newtheorem{corollary}[theorem]{Corollary}

\renewcommand{\qed}{\nobreak \ifvmode \relax \else
      \ifdim\lastskip<1.5em \hskip-\lastskip
      \hskip1.5em plus0em minus0.5em \fi \nobreak
      \vrule height0.75em width0.5em depth0.25em\fi}

\theoremstyle{definition}
\newtheorem{example}[theorem]{Example}
\newtheorem{definition}[theorem]{Definition}	

\centerline{{\LARGE A Multi-Dimensional Block-Circulant Perfect Array Construction}}
\medskip
\centerline{\large Samuel T. Blake}
\smallskip
\centerline{\large \it School of Mathematical Sciences, Monash University, Australia}
\bigskip

\begin{abstract}
\noindent{\sc Abstract.}  We present a $N$-dimensional generalization of the two-dimensional block-circulant 
perfect array construction by \cite{Blake2013}. As in \cite{Blake2013}, the families of $N$-dimensional arrays possess pairwise \textit{good} zero correlation zone (ZCZ) cross-correlation. Both constructions 
use a perfect autocorrelation sequence with the array orthogonality property (AOP). 
\end{abstract}

This paper presents a generalization of the 2-dimensional block-circulant 
array construction by \cite{Blake2013}. The generalized construction works in any number
of dimensions, but is limited to the same size in each dimension as the original two-dimensional
construction. Throughout the paper we follow the notation of \cite{Blake2013}. \\

We begin by stating the 2-dimensional construction for families of arrays with perfect 
autocorrelation and good cross-correlation as given in \cite{Blake2013}. 

\bigskip

\noindent\textbf{Construction I} \quad Let $\textbf{a}=[a_0,a_1,\cdots, a_{n-1}]$ be a perfect sequence with the AOP 
for the divisor, $d$, and and $\textbf{c} = [c(0), c(1), \cdots, c(d-1)]$ is a block of $d$ perfect 
sequences -- each of length $m$, where $m = 0 \mod d$. We construct a family of arrays, $\textbf{S}_k$, such that
\begin{center}
$\textbf{S}_k = \left[S_{i,j}\right]_k = a_j\, c(j \mod d)_{w \lfloor j/d \rfloor + k (j \mod d) + i}$
\end{center}
for $0 \leq i < n$, $0 \leq j < m$,  {\bf a} has the AOP for the divisor $d$, $0 < k \leq m$, and $w=m/d$.

\bigskip

This construction produces perfect arrays up to size $\mathit{r}^2 \times r^2$ over $r$ roots of unity. Each pair of distinct arrays has $d^2$ non-zero cross-correlation values, as $d \ll r$ we say the array has good
ZCZ cross-correlation. \\

The generalized $N$-dimensional construction is given as follows. 

\bigskip

\noindent\textbf{Construction II} \quad Let $\textbf{a}=[a_0,a_1,\cdots, a_{n-1}]$ be a perfect sequence 
 with the AOP for the divisor, $d$, and $\textbf{c} = [c(0), c(1), \cdots, c(d-1)]$ is a block of $d$ perfect 
sequences -- each of length $m$, where $m = 0 \mod d$. We construct a family of $k$ $N$-dimensional perfect 
arrays, $\textbf{S}_k$, such that $$\textbf{S}_k = \left[S_{i_0, i_1, \cdots, i_{N-2}, j}\right]_k = 
a_j\, \prod_{v=0}^{N-2} c(j \mod d)_{w \lfloor j/d \rfloor + k (j \mod d) + i_v}$$
for $0 \leq i_0, i_1, \cdots, i_{N-2} < n$, $0 \leq j < m$, $0 < k \leq m$, and $w = m/d$. 

\bigskip

Each distinct pair of arrays from the family has $d^2$ non-zero cross-correlation values, regardless of the 
size of $N$. Thus, the ratio of zero to non-zero cross-correlation values is larger for higher 
dimensional arrays. \\

We show that $\textbf{S}_k$ from Construction II has perfect autocorrelation ($k_1=k_2$) and $\textbf{S}_{k_1}$, 
$\textbf{S}_{k_2}$ has \textit{good} cross-correlation ($k_1 \neq k_2$). We now compute the 
cross-correlation of $\textbf{S}_{k_1}$ with $\textbf{S}_{k_2}$, ($0 < k_1,k_2 \leq m$) for 
shift $s_0, s_1, \cdots, s_{N-1}$:

\begin{align*}
&\theta_{\textbf{S}_{k_1}, \textbf{S}_{k_2}}(s_0, s_1, \cdots, s_{N-1})\\
&= \sum_{j=0}^{n-1} \sum_{i_0=0}^{m-1}\sum_{i_1=0}^{m-1} \cdots \sum_{i_{N-2}=0}^{m-1} 
	\left[S_{i_0,i_1, \cdots, i_{N-2},j}\right]_{k_1}\left[S_{i_0+s_0, i_1+s_1, \cdots, j+s_{N-1}}^*\right]_{k_2}\\
&= \sum_{j=0}^{n-1} \sum_{i_0=0}^{m-1}\sum_{i_1=0}^{m-1} \cdots \sum_{i_{N-2}=0}^{m-1} 
		\left(a_j\, \left(\prod_{v=0}^{N-2} c(j \mod d)_{w \lfloor j/d \rfloor + k_1 (j \mod d) + i_v}\right)
		a_{j+s_{N-1}}^* \times\right.\\ 
&\quad\quad\quad\quad \left.\left(\prod_{v=0}^{N-2} c(j+s_{N-1} \mod d)_{w \lfloor (j + s_{N-1})/d \rfloor + k_2 (j + s_{N-1} \mod d) + i_v + s_v}^*\right)\right)\\
&= \sum_{j=0}^{n-1}\left(a_j \, a_{j+s_{N-1}}^*  \sum_{i_0=0}^{m-1}\sum_{i_1=0}^{m-1} \cdots \sum_{i_{N-2}=0}^{m-1} 
\left(\left(\prod_{v=0}^{N-2} c(j \mod d)_{w \lfloor j/d \rfloor + k_1 (j \mod d) + i_v}\right)\times\right.\right.\\
&\quad\quad\quad\quad \left.\left.\left(\prod_{v=0}^{N-2} c(j+s_{N-1} \mod d)_{w \lfloor (j + s_{N-1})/d \rfloor + k_2 (j + s_{N-1} \mod d) + i_v + s_v}^*\right)\right)\right) \tag*{(7)}\\
\end{align*}
Consider the case when $s_{N-1} = 0 \mod  d$. Let $s_{N-1} = l \, d$ and perform the change of coordinates
$j = qd + r$, ($r<d$). Then the inner summation in (7) becomes
\begin{align*}
&\sum_{i_0=0}^{m-1}\sum_{i_1=0}^{m-1} \cdots \sum_{i_{N-2}=0}^{m-1} 
\left(\left(\prod_{v=0}^{N-2} c(r)_{k_1 r + i_v}\right)
\left(\prod_{v=0}^{N-2} c(r)_{lw + k_2 r + i_v + s_v}^*\right)\right)\\
&= \left(\sum_{i_0=0}^{m-1} c(r)_{k_1 r + i_0} \, c(r)_{lw + k_2 r + i_0 + s_0}^*\right)
\left(\sum_{i_1=0}^{m-1} c(r)_{k_1 r + i_1} \, c(r)_{lw + k_2 r + i_1 + s_1}^*\right) \times\\
&\qquad\left(\sum_{i_2=0}^{m-1} c(r)_{k_1 r + i_2} \, c(r)_{lw + k_2 r + i_2 + s_2}^*\right)\times \cdots \times 
\left(\sum_{i_{N-2}=0}^{m-1} c(r)_{k_1 r + i_{N-2}} \, c(r)_{lw + k_2 r + i_{N-2} + s_{N-2}}^*\right)
\end{align*}
which is independent of $q$. Thus, (7) becomes 
\begin{align*}
\sum_{r=0}^{d-1}
\left(\left(\sum_{q=0}^{n/d-1}a_{qd+r} \, a_{qd+r+s_{N-1}}^*\right)
\left(\sum_{i_0=0}^{m-1} c(r)_{k_1 r + i_0} \, c(r)_{lw + k_2 r + i_0 + s_0}^*\right)
\left(\sum_{i_1=0}^{m-1} c(r)_{k_1 r + i_1} \, c(r)_{lw + k_2 r + i_1 + s_1}^*\right)\right. \times\\
\left.\left(\sum_{i_2=0}^{m-1} c(r)_{k_1 r + i_2} \, c(r)_{lw + k_2 r + i_2 + s_2}^*\right)\times \cdots \times 
\left(\sum_{i_{N-2}=0}^{m-1} c(r)_{k_1 r + i_{N-2}} \, c(r)_{lw + k_2 r + i_{N-2} + s_{N-2}}^*\right)\right) \tag*{(8)}
\end{align*}
As \textbf{a} has the AOP for the divisor $d$,  the summation $\sum_{q=0}^{n/d-1}a_{qd+r} \, a_{qd+r+s_{N-1}}^* = 0$ 
for $s_{N-1} \neq 0 \mod n/d$, thus $\theta_{\textbf{S}_{k_1}, \textbf{S}_{k_2}}(s_0, s_1, \cdots, s_{N-1})=0$
for $s_{N-1} \neq 0 \mod n/d$. Otherwise, for $s_{N-1} = 0 \mod n/d$, (8) is zero when one of the remaining $N-2$ summations are zero.
The summation over $i_v$ is non-zero for $(k_2-k_1)r + s_{N-1} m/d^2 + s_v = 0 \mod m$. For each of the 
$s_{N-1} = 0, n/d, 2n/d, \cdots, (d-1)n/d$, there are $d$ solutions for $s_v$, one for each $r = 0, 1, \cdots, d-1$. Thus, there
are $d^2$ non-zero cross-correlation values for $s_{N-1} = 0 \mod d$.  Now consider the case
$k_1 = k_2$ (autocorrelation), (note that $s_{N-1} = 0 \mod n/d$), then (8) becomes 
\begin{align*}
\left(\sum_{r=0}^{d-1}\sum_{q=0}^{n/d-1}a_{qd+r} \, a_{qd+r+s_{N-1}}^*\right)
\left(\sum_{i_0=0}^{m-1} c(r)_{i_0} \, c(r)_{lw + i_0 + s_0}^*\right)
\left(\sum_{i_1=0}^{m-1} c(r)_{i_1} \, c(r)_{lw + i_1 + s_1}^*\right) \times\\
\left(\sum_{i_2=0}^{m-1} c(r)_{i_2} \, c(r)_{lw + i_2 + s_2}^*\right)\times \cdots \times 
\left(\sum_{i_{N-2}=0}^{m-1} c(r)_{i_{N-2}} \, c(r)_{lw + i_{N-2} + s_{N-2}}^*\right) \tag*{(8)}
\end{align*}
As \textbf{a} has the AOP for the divisor $d$, the left double summation is zero. Thus the autocorrelation is zero for $s_{N-1} = 0 \mod  d$.\\

Now, consider the case when $s_{N-1} \neq 0 \mod  d$. Let $s_{N-1} = l \, d + s$, ($s<d$), and perform the change of coordinates
$j = qd + r$, ($r<d$). Then (7) becomes
\begin{align*}
&\sum_{r=0}^{d-1} \sum_{q=0}^{n/d-1}\left(a_{qd+r} \, a_{qd+r+s_{N-1}}^*  \sum_{i_0=0}^{m-1}\sum_{i_1=0}^{m-1} \cdots \sum_{i_{N-2}=0}^{m-1} 
\left(\left(\prod_{v=0}^{N-2} c(r)_{q w + k_1 r + i_v}\right)\times\right.\right.\\
&\quad\quad\quad\quad \left.\left.\left(\prod_{v=0}^{N-2} c(r+s \mod d)_{q w + l w + w \lfloor (r+s)/d \rfloor + k_2 (r+s \mod d) + i_v + s_v}^*\right)\right)\right)\\
&= \sum_{r=0}^{d-1} \sum_{q=0}^{n/d-1}\left(a_{qd+r} \, a_{qd+r+s_{N-1}}^*
	\left(\sum_{i_0=0}^{m-1} c(r)_{q w + k_1 r + i_0} \, c(r+s \mod d)_{q w + l w + w \lfloor (r+s)/d \rfloor + k_2 (r+s \mod d) + i_0 + s_0}^*\right) \times \right.\\
	&\qquad \left(\sum_{i_1=0}^{m-1} c(r)_{q w + k_1 r + i_1} \, c(r+s \mod d)_{q w + l w + w \lfloor (r+s)/d \rfloor + k_2 (r+s \mod d) + i_1 + s_1}^*\right) \times\cdots\times \\
	&\qquad\left. \left(\sum_{i_{N-2}=0}^{m-1} c(r)_{q w + k_1 r + i_{N-2}} \, c(r+s \mod d)_{q w + l w + w \lfloor (r+s)/d \rfloor + k_2 (r+s \mod d) + i_{N-2} + s_{N-2}}^*\right)\right) \tag*{(9)}
\end{align*}
For each of the inner summations above, shifting both sequences $qw$ places does not change the cross-correlation. So (9) becomes
\begin{align*}
&= \sum_{r=0}^{d-1}\left( \left(\sum_{q=0}^{n/d-1}a_{qd+r} \, a_{qd+r+s_{N-1}}^*\right)
	\left(\sum_{i_0=0}^{m-1} c(r)_{k_1 r + i_0} \, c(r+s \mod d)_{l w + w \lfloor (r+s)/d \rfloor + k_2 (r+s \mod d) + i_0 + s_0}^*\right) \times \right.\\
	&\qquad \left(\sum_{i_1=0}^{m-1} c(r)_{k_1 r + i_1} \, c(r+s \mod d)_{l w + w \lfloor (r+s)/d \rfloor + k_2 (r+s \mod d) + i_1 + s_1}^*\right) \times\cdots\times \\
	&\qquad\left. \left(\sum_{i_{N-2}=0}^{m-1} c(r)_{k_1 r + i_{N-2}} \, c(r+s \mod d)_{l w + w \lfloor (r+s)/d \rfloor + k_2 (r+s \mod d) + i_{N-2} + s_{N-2}}^*\right)\right).
\end{align*}
As the sequence \textbf{a} has the AOP, the summation 
$$\sum_{q=0}^{n/d-1}a_{qd+r} \, a_{qd+r+s_{N-1}}^*=0$$
for all $r$ and $s_{N-1}$ such that $s_{N-1} \neq 0 \mod d$. Thus the cross-correlation of $\textbf{S}_{k_1}$ with 
$\textbf{S}_{k_2}$ is zero for $h \neq 0 \mod d$ and all $v$. So Construction II produces a family of $m$ 
perfect $k$-dimensional arrays with pairwise good cross-correlation. \\

It has been shown that perfect $N$-dimensional arrays can be constructed up to size $r^2\times r^2\times\cdots$ 
over $r$ roots of unity. An exponential number of such arrays exist for each size. Furthermore, families of arrays 
possess good cross-correlation properties. Importantly, despite the size of $N$, the number of non-zero 
cross-correlation values is always $d^2$, so the ratio of the number of non-zero cross-correlation values to the
number of entries in the arrays can be made as small as desired. 

\bibliographystyle{abbrv}

\section*{Appendix I -- Implementation of the Construction}
In this appendix we give a \href{http://www.wolfram.com}{Mathematica} (version 8.0)  implementation of the construction.
(Arrays are given in index notation, that is, the mapping: $e^{2 \pi \sqrt{-1} s_n/r} \rightarrow s_n$.)  \\

We begin with the code for periodic cross-correlation, {\tt XCV} and autocorrelation, {\tt ACV}:
{\small
\begin{verbatim}
In[1]:= XCV[a_, b_, r_Integer] := Block[{A, B},
  A = Developer`ToPackedArray[Exp[(2. Pi I a)/r]];
  B = Developer`ToPackedArray[Exp[(-2. Pi I b)/r]];
  Chop[ListCorrelate[A, B, 1], 1*^-5]]

In[2]:= ACV[m_, r_Integer] := XCV[m, m, r]
\end{verbatim}}
The $N$-dimensional array construction, {\tt ConstructionND}, takes as input the multiplying sequence,
\textbf{a}, the block of $d$ perfect sequences, \textbf{c}, the parameter, $k$, the number of 
dimensions, $dims$, and the number of roots of unity 
{\small
\begin{verbatim}
In[3]:= aref = #1[[#2 + 1]] &;

In[4]:= ConstructionND[a_, c_, k_, dims_] := 
 With[{d = Length[c], l = Length[First[c]]},
    Table[
       Array[
          indexND[a, c, k, d, l, j, ##] &, Table[l, {dims - 1}], 0], 
   {j, 0, Length[a] - 1}]]

In[5]:= indexND[a_, c_, k_, d_, l_, index__] := 
 With[{w = Length[First[c]]/d, j = First @ {index}},
     aref[a, j] + Product[aref[aref[c, Mod[j, d]], 
        Mod[w Floor[j/d] + k Mod[j, d] + i, l]], {i, Rest @ {index}}]]
\end{verbatim}}

We construct a perfect $4 \times 4 \times 4 \times 4$ binary array and compute its periodic autocorrelation: 
{\small
\begin{verbatim}
In[6]:= ConstructionND[FrankSequence[2], {#, Decimate[#, 3]} & @ FrankSequence[2], 0, 4];
Mod[%, 2]
Count[ACV[%, 2], n_?NumericQ /; n != 0., Infinity]
\end{verbatim}}
\leftline{\text{\tt Out[7]= }$\left[
\begin{array}{cccc}
 \left[
\begin{array}{cccc}
 0 & 0 & 0 & 1 \\
 0 & 0 & 0 & 1 \\
 0 & 0 & 0 & 1 \\
 1 & 1 & 1 & 0
\end{array}
\right] & \left[
\begin{array}{cccc}
 0 & 0 & 0 & 1 \\
 0 & 0 & 0 & 1 \\
 0 & 0 & 0 & 1 \\
 1 & 1 & 1 & 0
\end{array}
\right] & \left[
\begin{array}{cccc}
 0 & 0 & 0 & 1 \\
 0 & 0 & 0 & 1 \\
 0 & 0 & 0 & 1 \\
 1 & 1 & 1 & 0
\end{array}
\right] & \left[
\begin{array}{cccc}
 1 & 1 & 1 & 0 \\
 1 & 1 & 1 & 0 \\
 1 & 1 & 1 & 0 \\
 0 & 0 & 0 & 1
\end{array}
\right] \\
 \left[
\begin{array}{cccc}
 0 & 1 & 0 & 0 \\
 1 & 0 & 1 & 1 \\
 0 & 1 & 0 & 0 \\
 0 & 1 & 0 & 0
\end{array}
\right] & \left[
\begin{array}{cccc}
 1 & 0 & 1 & 1 \\
 0 & 1 & 0 & 0 \\
 1 & 0 & 1 & 1 \\
 1 & 0 & 1 & 1
\end{array}
\right] & \left[
\begin{array}{cccc}
 0 & 1 & 0 & 0 \\
 1 & 0 & 1 & 1 \\
 0 & 1 & 0 & 0 \\
 0 & 1 & 0 & 0
\end{array}
\right] & \left[
\begin{array}{cccc}
 0 & 1 & 0 & 0 \\
 1 & 0 & 1 & 1 \\
 0 & 1 & 0 & 0 \\
 0 & 1 & 0 & 0
\end{array}
\right] \\
 \left[
\begin{array}{cccc}
 0 & 1 & 0 & 0 \\
 1 & 0 & 1 & 1 \\
 0 & 1 & 0 & 0 \\
 0 & 1 & 0 & 0
\end{array}
\right] & \left[
\begin{array}{cccc}
 1 & 0 & 1 & 1 \\
 0 & 1 & 0 & 0 \\
 1 & 0 & 1 & 1 \\
 1 & 0 & 1 & 1
\end{array}
\right] & \left[
\begin{array}{cccc}
 0 & 1 & 0 & 0 \\
 1 & 0 & 1 & 1 \\
 0 & 1 & 0 & 0 \\
 0 & 1 & 0 & 0
\end{array}
\right] & \left[
\begin{array}{cccc}
 0 & 1 & 0 & 0 \\
 1 & 0 & 1 & 1 \\
 0 & 1 & 0 & 0 \\
 0 & 1 & 0 & 0
\end{array}
\right] \\
 \left[
\begin{array}{cccc}
 1 & 1 & 1 & 0 \\
 1 & 1 & 1 & 0 \\
 1 & 1 & 1 & 0 \\
 0 & 0 & 0 & 1
\end{array}
\right] & \left[
\begin{array}{cccc}
 1 & 1 & 1 & 0 \\
 1 & 1 & 1 & 0 \\
 1 & 1 & 1 & 0 \\
 0 & 0 & 0 & 1
\end{array}
\right] & \left[
\begin{array}{cccc}
 1 & 1 & 1 & 0 \\
 1 & 1 & 1 & 0 \\
 1 & 1 & 1 & 0 \\
 0 & 0 & 0 & 1
\end{array}
\right] & \left[
\begin{array}{cccc}
 0 & 0 & 0 & 1 \\
 0 & 0 & 0 & 1 \\
 0 & 0 & 0 & 1 \\
 1 & 1 & 1 & 0
\end{array}
\right]
\end{array}
\right]$}
\medskip
\leftline{\text{\tt Out[8]= }$1$}

\medskip

The following generalization of {\tt indexND} allows for computations over the quaternions:

{\small
\begin{verbatim}
In[9]:= NonCommutativeDot[e_] := e

In[10]:= indexND[a_, c_, k_, d_, l_, index__] /; ! FreeQ[a, Quaternion] := 
 With[{w = Length[First[c]]/d, j = First @ {index}},
  aref[a, j] ** Apply[NonCommutativeDot, 
    Table[aref[aref[c, Mod[j, d]], Mod[w Floor[j/d] + k Mod[j, d] + i, l]], {i, Rest@{index}}]]]
\end{verbatim}}
The following is a perfect quaternion sequence which posesses the array orthogonality property
for the divisor 4: 
$$[1,k ,1,-k ,-i,-k ,i,-k ,-1,k,-1,-k ,i,-k ,-i,-k]$$ 
We use this sequence to construct a perfect $16\times 16$ quaternion array.
{\small
\begin{verbatim}
In[11]:= q16 = {Quaternion[1, 0, 0, 0], Quaternion[0, 0, 0, 1], 
   Quaternion[1, 0, 0, 0], Quaternion[0, 0, 0, -1], 
   Quaternion[0, -1, 0, 0], Quaternion[0, 0, 0, -1], 
   Quaternion[0, 1, 0, 0], Quaternion[0, 0, 0, -1], 
   Quaternion[-1, 0, 0, 0], Quaternion[0, 0, 0, 1], 
   Quaternion[-1, 0, 0, 0], Quaternion[0, 0, 0, -1], 
   Quaternion[0, 1, 0, 0], Quaternion[0, 0, 0, -1], 
   Quaternion[0, -1, 0, 0], Quaternion[0, 0, 0, -1]};
   
In[12]:= ConstructionND[q16, {#, Decimate[#, 3], RotateRight[#, 2], #} & @ q16, 0, 2]
Count[QuaternionCV2D[%], n_?NumericQ /; n != 0., Infinity]
\end{verbatim}} 
\noindent$\text{\tt Out[12]= }\left[
\begin{array}{cccccccccccccccc}
 1 & k  & 1 & -k  & -i & -k  & i & -k  & -1 & k  & -1 & -k  & i & -k  & -i & -k  \\
 k  & 1 & j & -1 & j & 1 & k  & 1 & -k  & 1 & -j & -1 & -j & 1 & -k  & 1 \\
 -i & -k  & 1 & k  & 1 & -k  & -i & -k  & i & -k  & -1 & k  & -1 & -k  & i & -k  \\
 -k  & 1 & -k  & -1 & j & -1 & -j & -1 & k  & 1 & k  & -1 & -j & -1 & j & -1 \\
 -1 & -j & 1 & -j & i & j & i & -j & 1 & -j & -1 & -j & -i & j & -i & -j \\
 -j & -1 & -k  & -1 & k  & -1 & j & 1 & j & -1 & k  & -1 & -k  & -1 & -j & 1 \\
 i & j & 1 & j & -1 & j & -i & -j & -i & j & -1 & j & 1 & j & i & -j \\
 j & -1 & -j & -1 & k  & 1 & k  & -1 & -j & -1 & j & -1 & -k  & 1 & -k  & -1 \\
 1 & -k  & 1 & k  & -i & k  & i & k  & -1 & -k  & -1 & k  & i & k  & -i & k  \\
 -k  & 1 & -j & -1 & -j & 1 & -k  & 1 & k  & 1 & j & -1 & j & 1 & k  & 1 \\
 -i & k  & 1 & -k  & 1 & k  & -i & k  & i & k  & -1 & -k  & -1 & k  & i & k  \\
 k  & 1 & k  & -1 & -j & -1 & j & -1 & -k  & 1 & -k  & -1 & j & -1 & -j & -1 \\
 -1 & j & 1 & j & i & -j & i & j & 1 & j & -1 & j & -i & -j & -i & j \\
 j & -1 & k  & -1 & -k  & -1 & -j & 1 & -j & -1 & -k  & -1 & k  & -1 & j & 1 \\
 i & -j & 1 & -j & -1 & -j & -i & j & -i & -j & -1 & -j & 1 & -j & i & j \\
 -j & -1 & j & -1 & -k  & 1 & -k  & -1 & j & -1 & -j & -1 & k  & 1 & k  & -1
\end{array}
\right]$\\

\medskip

\noindent \text{\tt Out[13]= }$1$\\

We now construct a family of 9 perfect $9\times 9 \times 9 \times 9$ arrays 
over 3 roots of unity. The perfect Frank sequence: $[0, 0, 0, 0, 1, 2, 0, 2, 1]$, is 
used as the base sequence. 
{\small
\begin{verbatim}
In[14]:= family3D = Table[
   ConstructionND[FrankSequence[3], 
       {Decimate[#, 2], Decimate[#, 5], Decimate[#, 7]} & @ FrankSequence[3], n, 4], 
  {n, 9}];
\end{verbatim}} 
Each pair of sequences from the family have $d^2 = 3^2$ non-zero cross-correlation
values.  Here's the non-zero cross-correlation values of an example pair:
{\small
\begin{verbatim}
In[15]:= XCV[family3D[[1]], family3D[[2]], 3];
Cases[%, n_?NumericQ /; n != 0., Infinity]

Out[16]= {2187., 2187., 2187., -1093.5 - 1894. I, -1093.5 + 1894. I, 2187., -1093.5 + 1894. I, 
-1093.5 - 1894. I, 2187.}
\end{verbatim}} 
Here we generate all pairs of arrays from the family and count the number of non-zero 
cross-correlation values for each pair. 
{\small
\begin{verbatim}
In[17]:= Count[XCV[#1, #2, 3], n_?NumericQ /; n != 0., Infinity] & @@@ Tuples[family3D, 2]

Out[17]= {1, 9, 9, 9, 9, 9, 9, 9, 9, 9, 1, 9, 9, 9, 9, 9, 9, 9, 9, 9, 1, 9, 9, 9, 9, 9, 9, 9, 9, 
9, 1, 9, 9, 9, 9, 9, 9, 9, 9, 9, 1, 9, 9, 9, 9, 9, 9, 9, 9, 9, 1, 9, 9, 9, 9, 9, 9, 9, 9, 9, 1, 
9, 9, 9, 9, 9, 9, 9, 9, 9, 1, 9, 9, 9, 9, 9, 9, 9, 9, 9, 1}
\end{verbatim}} 
(The 1's above correspond to the autocorrelation of non distinct pairs.)

\end{document}